\documentstyle[a4,amstex]{article}

\newcommand{\cplx}{{\Bbb C}}
\newcommand{\zint}{{\Bbb Z}}

\newcommand{\ep}{\epsilon}
\newcommand{\luN}[1]{{\mbox{}}^N {#1}}
\newcommand{\lui}[1]{{\mbox{}}^{\infty} {#1}}
\newcommand{\slt}{\hat{{\frak s}{\frak l}}_2}
\newcommand{\sll}{{\frak s}{\frak l}_2}

\newcommand{\setN}{\{1,2,\dots,N \}}

\newcommand{\cV}{{\cal V}}

\newcommand{\BU }{{\Bbb U}}
\newcommand{\BY}{{\Bbb Y}}
\newcommand{\LJ}{L^{(J)}}
\newcommand{\LB}{L^{(B)}}
\newcommand{\Om}{\Omega_M^{(N)}}
\newcommand{\Omw}{\Omega_M^{(N)}(w)}
\newcommand{\Omm}{\Omega_M^{(N-2)}}

\newcommand{\FS}{(\cplx[w_1,\dots,w_N] \otimes (\otimes^N V))_F}
\renewcommand{\l}{\lambda}
\newcommand{\s}{\sigma}

\newcommand{\h}[1]{{#1}^{-1}}
\newcommand{\p}{\partial}
\newcommand{\be}{\begin{enumerate}}
\newcommand{\ee}{\end{enumerate}}
\newcommand{\ba}{\begin{array}}
\newcommand{\ea}{\end{array}}
\newcommand{\beq}{\begin{equation}}
\newcommand{\eeq}{\end{equation}}
\newcommand{\bqa}{\begin{eqnarray}}
\newcommand{\eqa}{\end{eqnarray}}
\newcommand{\bqas}{\begin{eqnarray*}}
\newcommand{\eqas}{\end{eqnarray*}}
\newcommand{\halmos}{\rule{5pt}{5pt}}

\newcommand{\iN}{i=1,\dots,N}

\numberwithin{equation}{section}

\begin{document}

\begin{titlepage}
\pagestyle{empty}
\begin{center}
\mbox{} \\
\vspace{3cm}
\begin{Large}
{\bf Semi-infinite wedges \\ and the conformal limit of the fermionic \\
Calogero-Sutherland Model with spin $\frac{1}{2}$  } \\

\end{Large}
\vspace{1.5cm}
{\large Denis Uglov} \\  Research Institute for Mathematical Sciences\\
Kyoto University, Kyoto 606, Japan \\e-mail: duglov@@kurims.kyoto-u.ac.jp \\ 
\vspace{1cm}
January 1996  \\
\vspace{2cm}
\begin{abstract}

The conformal limit over an anti-ferromagnetic vacuum of the fermionic spin  $\frac{1}{2}$ Calogero-Sutherland Model is derived by using the wedge product formalism. The space of states in the conformal limit is identified with the Fock space of two complex fermions, or, equivalently, with a tensor product of an irreducible level-1 module of $\slt$ and a Fock space module of the Heisenberg algebra.
 The Hamiltonian and the Yangian generators of the Calogero-Sutherland Model are represented in terms of  $\slt$ currents and bosons. At special values of the coupling constant they give rise to the Hamiltonian and the Yangian generators of the conformal limit of the Haldane-Shastry Model acting in  an irreducible level-1 module of $\slt$. At generic values of the coupling constant the space of states is decomposed into irreducible representations of the Yangian.

\end{abstract}

\end{center}

\end{titlepage}
\section{Introduction}

Solvable models with long-range interaction have become  a rapidly developing area of research over the course of the past few years. One important source of these new developments was the paper \cite{BGHP} where the original spinless Calogero-Sutherland Model \cite{CS} had been generalized so as to include particles with spin, and where a connection between these generalized Calogero-Sutherland Models and the Haldane-Shastry long-range interacting spin chain \cite{H,S} had been established.
 An essential feature of both generalized Calogero-Sutherland and the Haldane-Shastry Models, as emphasized in \cite{BGHP}, is the presence of the Yangian symmetry algebra which commutes with the Hamiltonian. This infinite-dimensional algebra of symmetries can be efficiently used to compute  spectra of excitations and, to some extent, dynamical correlation functions \cite{H2}.
 Remarkably, the Yangian was found to be an exact symmetry of the generalized Calogero-Sutherland and the (trigonometric) Haldane-Shastry Models even when the last two are considered in finite volume and at fixed finite number of particles. This distinguishes the long-range interacting models from the class of solvable models with point interactions, such as the Heisenberg $XXX$ spin chain and the Hubbard Model where Yangian symmetry is exact only in thermodynamic limit taken over anti-ferromagnetic vacuum.       

The work on conformal limit of the spin-$\frac{1}{2}$ Haldane-Shastry Model, which was initiated in \cite{HHTBP} and continued in \cite{BPS} and \cite{BLS}, brought equally remarkable results. It was found, that in the conformal limit taken in the vicinity of the anti-ferromagnetic ground state, the space of states of the Haldane-Shastry Model can be identified with the direct sum of the two integrable, irreducible level-1 representations of the algebra $\slt$, so that the generators of the Yangian symmetry and the Hamiltonian are expressed in terms of $\slt$ currents.
 The decomposition of the space of states into irreducible representations of the Yangian provides a new basis of the level-1 representations of $\slt$ and, therefore, new character formulas for these representations. 
This new basis is written in terms of the Vertex Operators associated with the level-1 $\slt$-representations, which, in the context of the Haldane-Shastry Model are interpreted as creation operators of spinon excitations -- particles with spin $\frac{1}{2}$ and half-integer statistics \cite{H2}. These results were further generalized to include ${\frak s}{\frak l}_n $ Haldane-Shastry Model , and the $q$-deformed situation \cite{JKKMP}.   

In the case of finite number of particles several authors observed \cite{TH},\cite{Poly} , that the Haldane-Shastry spin chain can be considered a special subcase of the generalized Calogero-Sutherland Model where coordinates of the particles are frozen so that only the spin degrees of freedom remain relevant. 
This suggests, that a similar situation may be encountered in the conformal limit as well. In fact, this is a point of view adopted in \cite{BPS}.
 Thus the natural problem arises -- to find what is the conformal limit of the Calogero-Sutherland Model with spin and how to derive the Hamiltonian and the Yangian symmetry of the Haldane-Shastry Model from the corresponding Calogero-Sutherland objects. This is the main issue  with which we deal in this paper. Methodologically, a new feature of the present work is an extensive use of the wedge product formalism. About this formalism a reader can consult the works \cite{St},\cite{KMS} where it is introduced in the general, $q$-deformed setting.

The approach which we use can be summarized as follows. First we reformulate the finite-particle fermionic Calogero-Sutherland Model with spin $\frac{1}{2}$ in terms of the finite wedge product of infinite-dimensional spaces $V(z)$ = $ \cplx[z,z^{-1}] \otimes \cplx^2 $. In the space $V(z)$ and in the wedge product one naturally defines a level-0 action of $\slt$ . 

In the wedge product language  anti-ferromagnetic vacua of the Model have simple and explicit form. Taking advantage of this one can  implement  a conformal limit in the vicinity of any of these vacua by going from finite to semi-infinite wedge product. The wedge product formulation makes the transition to the conformal limit especially transparent.
 The expressions for the spin Calogero-Sutherland Hamiltonian and the Yangian generators remain well-defined when the finite wedge product is replaced with the infinte one. Next, the semi-infinite wedge product can be identified with a fixed-charge subspace of the Fock space of two complex fermions , the charge being defined by the vacuum around which one takes the conformal limit. 

Thus one derives a  Yangian action in the Fock space such, that it  commutes with the Calogero-Sutherland Hamiltonian defined in the same space. The semi-infinite wedge product is a level-1 reducible highest-weight representation of the algebra $\slt$. However, it is an irreducible representation of the direct sum of $\slt$ and  Heisenberg algebra $H$ commuting with $\slt$  \cite{KMS}; furthermore , it  can be decomposed  into  a tensor product of a level-1 irreducible representation of $\slt$ and the  boson Fock space generated by $H$ \cite{KMS}.
 The last decomposition has the meaning of  spin-charge separation of the space of states  in the conformal limit \cite{BPS}. The generators of $\slt$ assume the role of spin degrees of freedom, while the bosons forming the Heisenberg algebra -- the role of charge degrees of freedom. 

At two special values of the coupling constant of the Calogero-Sutherland Model the charge part of the space of states represented by the Fock space of bosons can be projected out, so that the property of the Yangian invariance remains intact. This gives the Hamiltonian and the Yangian generators of the Haldane-Shastry Model acting in one of the irreducible level-1 representations of $\slt$. 

On the other hand at the same two values of the coupling constant one can project out the spin degrees of freedom represented by the $\slt$-generators. This gives a third-degree Hamiltonian in the generators of the Heisenberg algebra. This Hamiltonian can be interpreted as the spinless Calogero-Sutherland Hamiltonian at the special value of the coupling constant, and can be compared to the collective-field spinless Calogero-Sutherland Hamiltonian recently introduced in \cite{Awata}. 

Finally at  generic, specifically not rational, values of the coupling constant in the fermionic spin Calogero-Sutherland Hamiltonian one can decompose the fermion Fock space into irreducible representations of the Yangian -- eigenspaces of the  Hamiltonian. 
To carry out this decomposition we use the ``fermion basis'',  named so by analogy with the spinon basis \cite{BPS}, which is formed by acting with creation operators of holes on a vacuum vector.
 The action of the Hamiltonian and the Yangian generators on this basis is expressed by means of the finite-particle fermionic Calogero-Sutherland Hamiltonian and Yangian generators -- the same objects from which we start, and then go on to take the conformal limit. However the  coupling constant is now different -- there is a finite renormalization by the amount equal to the dual Coxeter number of $\sll$.    

Recently in the paper \cite{Awata1} the collective-field description of spin Calogero-Sutherland Models was given. In this description the Hamiltonians are expressed in terms of several bosonic fields. It would be interesting to find a relationship between the Hamiltonian which we derive in this present paper and those of \cite{Awata1}.                

\mbox{}

The paper is organized as follows. In  section 1 we describe the finite wedge product and define in this product the action of fermionic Calogero-Sutherland Hamiltonian of spin $\frac{1}{2}$, as well as the actions  of  several algebras, one of which is the Yangian commuting with the Hamiltonian.
 In section 2 we consider the conformal limit and define the Calogero-Sutherland Model in the Fock space of complex fermions. In section 3 we describe the irreducible decomposition of the Fock space with respect to the Yangian action.

\section{Finite wedge product and fermionic Calogero-Sutherland Model 
with spin $\frac{1}{2}$}

The main aim of this section is to describe the fermionic Calogero-Sutherland Model of spin $\frac{1}{2}$ and its Yangian symmetry  \cite{BGHP} in the language of the wedge product formalism \cite{S,KMS}. In this section the number of particles in the Model is considered to be finite and equal to the number of factors in the wedge product, the last being identified with the space of states of the Model. 
Apart from the actions of the Calogero-Sutherland Hamiltonian and the Yangian, in the wedge product one can define actions of three Lie algebras: $\slt$, Heisenberg algebra and Virasoro algebra. None of these algebras commute with the Hamiltonian, and in the situation of finite number of particles neither of them has any apparent use as far as the Calogero-Sutherland Model is concerned. 
These Lie algebras, however, assume the central role in the conformal limit  where they become responsible for organization of the space of states, and where both the Hamiltonian and the Yangian are expressed in terms of their generators.

\subsection{Preliminaries}

The algebra $\slt $ is generated by elements $ \{ J_m^a , k , D \} $ where  $ m \in \zint $ , $ a = 1,2,3 $ and $ k $ and $ D $ are the central element and the degree operator respectively. In the normalization adopted in this paper the commutation relations of $ \slt $ have the following form:  
\begin{equation}
[ J_m^a , J_n^b]\; = \; 2i\ep^{abc}J_{m+n}^c \; + \; 2 k m \delta^{ab}\delta_{m+n,0}\quad , \quad [D,J_m^a] = m J_m^a \; . 
\end{equation}
The completely antisymmetric symbol $ \ep^{abc} $ which appears above is normalized in the usual way: $ \ep^{123} = 1 $ . 

Let $V = \cplx^2 $,  with basis $ \{ v_1 , v_2 \} $ and $ V(z) = \cplx [z,\h z ] \otimes V $, with basis $ \{ z^a v_{\epsilon} \} $; where  $ a \in \zint   $ , and $ \ep = 1 , 2 $. The space $V$ carries a spin 1/2 representation of $\sll $ given by the Pauli operators  $\{\s^a \}$, $ a = 1,2,3 $ :
\begin{equation}
\s^1 v_{\ep} = v_{3 - \ep}\quad ,\quad   \s^2 v_{\ep} = (-1)^{\ep + 1 } i v_{3 - \ep}\quad  ,\quad \s^3 v_{\ep }  = (-1)^{\ep + 1} v_{\ep} \quad ; \quad   \ep = 1 , 2 . 
\end{equation}
 The space $ V(z) $ carries  a zero-level infinite-dimensional representation of algebra $ \slt $. The generators $ \{ J_m^a  \} $, $ m\in \zint $ , $ a = 1,2,3 $ and $D$ of $\slt $ act in $ V(z) $  in the following way:          
\begin{equation}
J_m^a \quad = \quad   z^m \s^a \quad , \quad  D \quad = \quad  z\frac{\p}{\p z}\quad . \label{e: affsl21}
\end{equation}
Sometimes it will be convenient to use another notation for the basis vectors of $ V(z) $ : denote $ u_{\ep - 2 n } =  z^n v_{\epsilon}  $ ; $ n \in \zint $, $ \ep =1,2 $ . The vectors $ \{ u_k \} $ , $ k \in \zint $ then form a basis in $ V(z) $ . 

Now let us form a tensor product $ \otimes^N V(z) $  of $N$ spaces $ V(z) $ . This product is naturally identified with the space $ \cplx[z_1,\h z_1, z_2, \h z_2 , \dots , z_N , \h z_N ] \otimes ( \otimes^N V ) $. In this space one can define actions of several algebras. First of all, of course, there is  an action of $\slt $ given by the tensor product of  the expressions (\ref{e: affsl21}):
\begin{equation}    
J_m^a \quad = \quad \luN{J}_m^a = \sum_{i=1}^N z_i^m \s^a_i\quad, \quad D \quad = \quad \luN{D} = \sum_{i=1}^N D_i \; .  \label{e: affsl2N} 
\end{equation}
Above and throughout the paper the usual convention about lower indices is used : if an operator $ A $ is defined in a vector space $ \cV $ ( $ \cV \otimes \cV $ ) , then $ A_i $ ($ A_{ij})$ refers to the operator which acts trivially on all factors in a tensor product of several spaces $ \cV $ except the $ i $-th ( $i$-th and $j$-th ) factor(s) where it acts as $ A $. 

Besides $\slt$ in the tensor product  $ \otimes^N V(z) $ act other algebras. The first of these is an infinite-dimensional {\it Heisenberg algebra } $H$ with zero central element. It is generated by the elements $ \{ B_m \} $ , $ m \in \zint $ defined as the power-sums:  
\begin{equation}
 B_m \quad = \quad  \luN{B}_m  = \sum_{i=1}^N z_i^m  \; . \label{e: BBN}
\end{equation}
The significance of this algebra will become clear when we go over from the finite tensor product of the spaces $V(z)$ to (a wedge subquotient of ) a semi-infinite one. Here it suffices to notice that $ H $ commutes with the action of $\slt$. 

Next, one defines an action of the {\it Virasoro algebra }  $Vir$ with zero central charge by:
\begin{equation}   
L_m \quad = \quad  \luN{L}_m = -\sum_{i=1}^N z_i^m D_i \; , \quad  c = 0 \;, \quad m \in \zint. \label{e: VirN}
\end{equation}   
The commutation relations between the generators of $Vir$ and $\slt$ resemble those encountered in the Sugawara construction:
\begin{equation}   
   [ \luN{L}_m , \luN{J}_n^a ] \quad = - n\;\; \luN{J}_{n+m}^a \; .  \label{e: LJN}
\end{equation}
Whereas the generators of $Vir$ and $H$ obey a similar type of relations, which is reminiscent of the commutation relations between the free boson Virasoro algebra and creation/annihilation operators of bosons \cite{KR}: 
\begin{equation}  
[ \luN{L}_m , \luN{B}_n^a ] \quad = - n\;\; \luN{B}_{n+m}^a \; . \label{e: LBN}
\end{equation}

Now we come to a definition of several algebras which are associated with the long-range interacting solvable models such as the Calogero-Sutehrland Models with spin -- alteratively called  Dynamical long-range Models  \cite{BGHP}, and the Haldane-Shastry Model \cite{H,H2},\cite{S}.  

In the tensor product $\otimes^N V(z) $ =  $ \cplx[z_1,\h z_1, z_2, \h z_2 , \dots , z_N , \h z_N ] \otimes ( \otimes^N V ) $ define the coordinate  permutation operators $\{ K_{ij} \} $ :
\begin{equation*}  
K_{ij}z_i = z_j K_{ij}\quad,\qquad  [K_{ij},z_k] = 0 \quad,\quad k \not= i,j \;;
\end{equation*}
and the {\it Dunkl operators } \cite{Dunkl,P} $  \{ d_i(\alpha) \}$ , $ \iN $ : 
\begin{equation}
d_i(\alpha) = \alpha D_i - i + \sum_{j > i}\theta_{ji}(K_{ij} - 1) - \sum_{j < i}\theta_{ij}(K_{ij} - 1) \; , \quad \alpha \in \cplx , \label{e: Dunkl} 
\end{equation}
where $ \theta_{ij} = z_i/(z_i - z_j) $.

Together, the operators $ \{ d_i(\alpha) , K_{ij}  \}$ satisfy the relations of the {\it degenerate Affine Hecke Algebra } :
\begin{gather*}
K_{ii+1}K_{i+1i+2}K_{ii+1} = K_{i+1i+2}K_{ii+1}K_{i+1i+2} \; ; \;  [K_{ii+1},K_{jj+1}] = 0 \;, \;  |i-j| \geq 2 , \\
K_{ii+1} d_i(\alpha) - d_{i+1}(\alpha) K_{ii+1} = 1 \; ; \;   [K_{ii+1},d_j(\alpha)] = 0 \;,\;  |i-j| \geq 2  , \\ 
    \mbox{}   [d_i(\alpha), d_j(\alpha)] = 0 \; .   \label{e: Hecke}
\end{gather*}

Starting from the Dunkl operators one defines, following \cite{BGHP} a family of mutually commuting quantities $\{ h^{(n)} \} $, $ n = 1,\dots, N $ :
\begin{equation*}
h^{(n)} (\alpha)= \sum_{i=1}^{N} d_i(\alpha) ^n \; .  
\end{equation*}
In this article only two of these quantities will be further considered: the $ h^{(1)}(\alpha) $ which is related to the energy operator of the Virasoro algebra (\ref{e: VirN}): 
\begin{equation*}
h^{(1)}(\alpha) =  -\alpha \; \; \luN{L}_0  - \frac{1}{2} N(N+1) \; ; 
\end{equation*}
and $ h^{(2)}(\alpha) $ which is related to the generalized Calogero-Sutherland Hamiltonian $ \luN{\hat{H}}_{CS}(\alpha) $ \cite{BGHP,P}: 
\begin{eqnarray}
 h^{(2)}(\alpha) & = & \alpha \; \; \luN{\hat{H}}_{CS}(\alpha) + 2\alpha (N+1)\; \; \luN{L}_0 + \frac{N(N+1)(2N+1)}{6} \; ; \nonumber \\  
  &    &  \label{e: HCSN}   \\ 
\luN{\hat{H}}_{CS}(\alpha) & = & \alpha \sum_{i=1}^{N} D_i^2  + 2\sum_{i=1}^{N}i D_i + 2\sum_{1\leq i < j \leq N} \theta_{ij}\left(D_i - D_j - \theta_{ji}(K_{ij} - 1)\right) \; .  \nonumber
\end{eqnarray}

The Dunkl operators, finally, give rise to a non-abelian symmetry algebra of the solvable hierarchy defined by $ \{ h^{(n)}(\alpha)\} $ and, consequently, a symmetry algebra of the Hamiltonian $\luN{\hat{H}}_{CS}(\alpha) $ \cite{BGHP}. This algebra is the {\it Yangian } $ Y(\sll) $ \cite{Drinfel'd}. It is defined by the six generators $ \{ \hat{Q}_0^a ,\hat{Q}_1^a \} $, $ a = 1,2,3 $: 
\begin{eqnarray}       
\hat{Q}_0^a \; = \;\luN{\hat{Q}}_0^a & = & \luN{J}_0^a \; = \; \sum_{i=1}^{N} \s_i^a  \quad ,  \label{e: Yangian0N}\\
\hat{Q}_1^a \; = \;\luN{\hat{Q}}_1^a & = &  \sum_{i=1}^{N} d_i(\alpha)\s_i^a  + \frac{i}{2} \sum_{1\leq i < j \leq N}\ep^{abc} \s_i^b \s_j^c \quad .\label{e: Yangian1N} 
\end{eqnarray}

It may be remarked, that the action of $ Y(\sll) $ and of the Hamiltonian $ \luN{\hat{H}}_{CS}(\alpha) $ can be restricted from $ \cplx[z_1,\h z_1, z_2, \h z_2 , \dots , z_N , \h z_N ] \otimes ( \otimes^N V ) $ to $ \cplx[z_1, z_2,  \dots , z_N ] \otimes ( \otimes^N V ) $ due to the fact that the Dunkl operators preserve the space of polynomials in $ z_1,\dots,z_N $. This observation will be important in the sec. 3 of this article.

The four algebras whose action in $ \otimes^N V(z) $ was introduced above: the $\slt $, the Heisenberg algebra $ H $, the Virasoro algebra $ Vir $ and the Yangian $ Y(\sll) $;  as well as the generalized Calogero-Sutherland Hamiltonian $ \luN{\hat{H}}_{CS}(\alpha) $ are the main objects that shall be considered separately and in relation to each other in subsequent sections. 

\subsection{Finite wedge product}

In order to introduce, following \cite{KMS}, the wedge product $ \wedge ^N V(z) $ of spaces $ V(z)$ one utilizes the operators of coordinate permutation $ K_{ij} $ described in the previous section, and operators of spin permutation $ P_{ij} $. The operator $ P_{ij}\; ,\; 1\leq i \not= j \leq N  $ acts in $ \otimes ^N V(z) $ = $ \cplx[z_1,\h z_1, z_2, \h z_2 , \dots , z_N , \h z_N ] \otimes ( \otimes^N V ) $ by exchanging $i$ -th and $j$-th factors in  $ \otimes^N V $. Define  $ \Omega $ $  \subset $  $ \otimes ^N V(z) $ as: \begin{equation*}  
      \Omega \quad = \quad \bigcup_{1\leq i \leq N-1} Ker( K_{ii+1} - P_{ii+1}) \; . 
\end{equation*}
The wedge product $ \wedge ^N V(z) $ is then defined as a quotient of the tensor product over the subspace $\Omega$:
\begin{equation} 
        \wedge ^N V(z) = \otimes ^N V(z) / \Omega \; \label{e: wedge1}  
\end{equation}
A vector $ u_{k_1} \wedge u_{k_2}\wedge \dots \wedge u_{k_N}  $  $\in $ $ \wedge ^N V(z) $  called {\it a wedge } thereafter, is defined as the image of the pure tensor $ u_{k_1} \otimes u_{k_2}\otimes \dots \otimes u_{k_N} $ under the quotient map (\ref{e: wedge1}). 

A wedge $w$:
\begin{equation*}
w = u_{k_1} \wedge u_{k_2}\wedge \dots \wedge u_{k_N} 
\end{equation*}
is called an {\it ordered wedge } provided $ k_1 > k_2 > \dots > k_N $. Thanks to the antisymmetry relation:
\begin{equation} 
      u_k \wedge u_l = - u_l \wedge u_k \; ,  \label{e: anticomm}
\end{equation}
which follows from (\ref{e: wedge1}), ordered wedges form a basis in the space  $ \wedge ^N V(z) $.

The important property of the subspace $ \Omega $ is that the actions of the algebras $ \slt $, $ H$, $ Vir$ and $ Y(\sll) $ {\it preserve } this subspace. Therefore these actions factor through the quotient map (\ref{e: wedge1}) and define actions of respective algebras in the wedge product  $ \wedge ^N V(z) $. The Hamiltonian $\luN{\hat{H}}_{CS}(\alpha) $ preserves $\Omega$ as well. 
So all the objects considered in the previous subsection ( with the obvious exception of the degenerate Affine Hecke Algebra)  carry over into the space $ \wedge ^N V(z) $ with their respective commutation relations intact. In particular (\ref{e: HCSN}) and (\ref{e: Yangian0N},\ref{e: Yangian1N}) define in $ \wedge ^N V(z) $ the operator $\luN{H}_{CS}(\alpha)$ and operators $\{\luN{ Q}_0^a ,\luN{Q}_1^a \} $, $ a = 1,2,3 $ ,  such that:   
\begin{equation*}
           [\luN{H}_{CS}(\alpha) \;,\;\luN{Q}_r^a  ]\; = \; 0 \quad , \quad r=0,1 \; ,
\end{equation*}
and $\{\luN{ Q}_0^a ,\luN{Q}_1^a \} $ satisfy the defining relations of the Yangian.
These operators can be identified with the fermionic spin-$ \frac{1}{2} $ Calogero-Sutherland Hamiltonian  and generators of the associated Yangian symmetry \cite{BGHP,P}. 

To clarify this, consider  $ {\cal F} $ $\subset$ $\otimes^N V(z) $:  
\begin{equation}
  {\cal F}  = \bigcap_{1\leq i \leq N-1} Ker( K_{ii+1} + P_{ii+1}) \; . \label{e: F}
\end{equation}
This is the subset of ``fermionic wave functions `` i.e. totally antisymmetric vectors in $ \cplx[z_1,\h z_1, z_2, \h z_2 , \dots , z_N , \h z_N ] \otimes ( \otimes^N V ) $ = $\otimes^N V(z) $. In virtue of the relation:
\begin{equation}  
\otimes^N V(z)  = \Omega \oplus {\cal F} \; , \label{e: Om+F}   
\end{equation}
the subspace ${\cal F}$ can be identified with $\wedge^N V(z) $. According to \cite{BGHP,P} the  fermionic spin-$ \frac{1}{2} $ Calogero-Sutherland Hamiltonian  and generators of the associated Yangian symmetry are defined by restriction on ${\cal F}$ of the Hamiltonian $\luN{\hat{H}}_{CS}(\alpha) $ (\ref{e: HCSN}) and the Yangian generators $\{\luN{\hat{Q}}_{0,F}^a ,\luN{\hat{Q}}_{1,F}^a \} $:     
\begin{eqnarray}       
\luN{\hat{Q}}_{0,F}^a & = & \luN{J}_0^a \; = \; \sum_{i=1}^{N} \s_i^a  \quad ,  \label{e: Yangian0NF}\\
\luN{\hat{Q}}_{1,F}^a & = &  \sum_{i=1}^{N} d_i(\alpha)\s_i^a  - \frac{i}{2} \sum_{1\leq i < j \leq N}\ep^{abc} \s_i^b \s_j^c \quad .\label{e: Yangian1NF} 
\end{eqnarray}
These generators coincide with (\ref{e: Yangian0N},\ref{e: Yangian1N}) up to the difference in sign in front of the double sum term in $\luN{\hat{Q}}_{1,F}^a$. 
The operators $\luN{\hat{H}}_{CS}(\alpha) $ and $\{\luN{\hat{Q}}_{0,F}^a ,\luN{\hat{Q}}_{1,F}^a \} $ preserve ${\cal F}$ and, therefore, the restriction of these operators on ${\cal F}$ is consistent with the Yangian defining relations and the commutativity between the Yangian and the Calogero-Sutherland Hamiltonian. 
Acting in ${\cal F}$ one can eliminate  the operators of coordinate permutation $K_{ij}$, replacing them with the operators $P_{ij}$ of spin permutation, by carrying $K_{ij}$ to the {\it right } of an expression and replacing $K_{ij}$ with $-P_{ij}$ in accordance with (\ref{e: F}). 
Let  $\luN{H}_{CS}(\alpha)_F$ and $\{\luN{Q}_{0,F}^a ,\luN{Q}_{1,F}^a \}$  be obtained from $ \luN{\hat{H}}_{CS}(\alpha) $ and $\{\luN{\hat{Q}}_{0,F}^a ,\luN{\hat{Q}}_{1,F}^a \} $ by carrying out this replacement. The expressions for these operators are: 
\begin{eqnarray}   
& & \label{e: HCSFN} \\
\luN{H}_{CS}(\alpha)_F & = & \sum_{i=1}^{N}( \alpha D_i^2 + (N+1) D_i) + \sum_{1\leq i \not= j \leq N} \theta_{ij}\left(D_i - D_j + \theta_{ji}(P_{ij} + 1)\right) , \nonumber  \\ 
& & \label{e: Yangian0FN} \\
Q_{0,F}^a &  =  & \luN{J}_0^a \; = \; \sum_{i=1}^{N} \s_i^a  \quad , \nonumber \\
& & \label{e: Yangian1FN} \\
Q_{1,F}^a  & = & \sum_{i=1}^{N}(\alpha D_i \s_i^a - N\s_i^a)   - \frac{i}{2} \sum_{1\leq i \not= j \leq N}\theta_{ij}\ep^{abc} \s_i^b \s_j^c  + \sum_{1\leq i \not= j \leq N}\theta_{ij} \s^a_i\: . \nonumber
\end{eqnarray}

Turning to the wedge product $\wedge^N V(z)$, which is identified with ${\cal F}$ as shown by (\ref{e: wedge1},\ref{e: Om+F}), one  can as well obtain a representation for $\luN{H}_{CS}(\alpha)$ and $\{ \luN{Q}_{0}^a ,\luN{Q}_{1}^a \} $ in which all the operators $K_{ij}$ are eliminated. In this case to eliminate a $K_{ij}$ one must carry it to the {\it left } of an expression and replace it then by $ -P_{ij}$. This follows from the relation:
\begin{equation*}    
Ker(K_{ii+1} - P_{ii+1}) = Im(K_{ii+1} + P_{ii+1})\;,\quad 1\leq i \leq N-1 \; ,\end{equation*}
and the definition of the wedge product (\ref{e: wedge1}). A straightforward computation shows, that $\luN{H}_{CS}(\alpha)$ and $\{ \luN{Q}_{0}^a ,\luN{Q}_{1}^a \} $ obtained from  $\luN{\hat{H}}_{CS}(\alpha)$ and $\{ \luN{\hat{Q}}_{0}^a ,\luN{\hat{Q}}_{1}^a \} $ by this $K_{ij}$ to $-P_{ij}$ replacement  coincide with the fermionic operators $\luN{H}_{CS}(\alpha)_F$ and $\{\luN{Q}_{0,F}^a ,\luN{Q}_{1,F}^a \}$  (\ref{e: HCSFN}-\ref{e: Yangian1FN}).
 Thus the wedge product can be considered as an alternative way to introduce the fermionic Calogero-Sutherland Model and its Yangian symmetry. 

One advantage of working with wedges is that some of the eigenvectors of the fermionic Calogero-Sutherland Hamiltonian have a very simple form when written in the wedge language. Let $M$ be an integer such, that $M$ and $N$ have the same parity. Then as can be checked by using formulas (\ref{e: HNw}) and (\ref{e: YNw}); the vectors:
\begin{equation}
|M\rangle ^{(N)} = u_M\wedge u_{M-1}\wedge u_{M-2}\wedge \dots \wedge u_{M-N+1}\; , \label{e: Nvac}
\end{equation}
are eigenvectors of the Hamiltonian and highest-weight vectors of the Yangian.
When both $M$ and $N$ are even, $|M \rangle ^{(N)}$ is a one-dimensional representation of the Yangian and represents an antiferromagnetic vacuum of the Model. When both $M$ and $N$ are odd , the vectors  $|M \rangle ^{(N)}$ and $ u_{M+1}\wedge |M-1\rangle ^{(N-1)} $ form a two-dimensional representation of the Yangian. This is a doubly-degenerate antiferromagnetic vacuum.

This indicates that the wedge-product description is suitable if we want to consider a conformal limit $N \rightarrow \infty$ over one of the antiferromagnetic vacua of the Model. This is indeed the case  as we shall try to demonstrate in the next section.    

For the later use let us write explicit formulas for  the action of $\luN{H}_{CS}(\alpha)$ and $\luN{Q}_{1}^a$ on an  ordered wedge $w$: 
\begin{multline*}
\text{Put:} \qquad \qquad w = u_{k_1} \wedge u_{k_2}\wedge \dots \wedge u_{k_N} = \\ 
z^{m_1}v_{\ep_1} \wedge z^{m_2}v_{\ep_2}\wedge \dots \wedge z^{m_N}v_{\ep_N} \;,\qquad k_i = \ep_i - 2 m_i \,, \\ 
k_1 > k_2 > \dots > k_N  \Rightarrow m_1 \leq m_2 \leq \dots \leq m_N \,.
\end{multline*}
Then the Hamiltonian acts on $w$ as follows:
\begin{eqnarray}
& & \label{e: HNw}\\
\luN{H}_{CS}(\alpha).w & = & \sum_{i=1}^{N}( \alpha m_i^2 + 2i m_i) w + 2\sum_{1\leq i < j \leq N} h_{ij}. w  \; , \nonumber \\ & & \text{where for $m \leq n$ } \quad  
h_{ij}. (\dots \wedge \underset{i}{z}^m v_{\ep} \wedge \dots \wedge \underset{j}{z}^n v_{\ep'}\wedge \dots  ) = \nonumber \\ & = & \sum_{r=1}^{n-m -1} (n-m-r) (\dots \wedge \underset{i}{z}^{m+r} v_{\ep} \wedge \dots \wedge \underset{j}{z}^{n-r} v_{\ep'} \wedge \dots )\; . \nonumber  
\end{eqnarray}
And for the Yangian action one has:
\begin{eqnarray}
& & \label{e: YNw} \\
& & \luN{Q}_{0}^a .w = \sum_{i=1}^{N} \s_i^a.w \;, \nonumber \\
& & \luN{Q}_{1}^a .w = \sum_{i=1}^{N}( \alpha m_i \s_i^a  -i \s^a_i) w +  \nonumber \\ 
& & + \sum_{i=1}^{N-1}\delta_{m_i,m_{i+1}}(\s_i^a - \s_{i+1}^a). w + \frac{i}{2} \sum_{1\leq i < j \leq N}\ep^{abc} \s_i^b \s_j^c.w  -  
\sum_{1\leq i < j \leq N} q^a_{ij}. w  \; , \nonumber\\
& & q^a_{ij}. (\dots \wedge \underset{i}{z}^m v_{\ep} \wedge \dots \wedge \underset{j}{z}^n v_{\ep'}\wedge \dots  ) = \nonumber \\ & & = \sum_{r=1}^{n-m -1} (\s_i^a - \s_j^a)(\dots \wedge \underset{i}{z}^{m+r} v_{\ep} \wedge \dots \wedge \underset{j}{z}^{n-r} v_{\ep'} \wedge \dots )\;, \quad  m \leq n . \nonumber  
\end{eqnarray}
Above and elsewhere $\s_i^a.w $ is defined as:
\begin{equation*}
\s_i^a.w = ( \dots \wedge z^{m_i} (\s^a v_{\ep_i}) \wedge \dots ) \; .
\end{equation*}
The formulas (\ref{e: HNw},\ref{e: YNw}) give the action of the Hamiltonian and the Yangian in the form which is suitable for a transition to the conformal limit in the neighborhood of one of the anti-ferromagnetic vacua (\ref{e: Nvac}) of the model.

\section{Semi-infinite wedge product and the conformal limit of
 the Calogero-Sutherland Model}

In this section we start by giving  a description of the semi-infinite wedge product following the definition given in \cite{KMS}. In the language of the Calogero-Sutherland Model this definition can be understood as follows. First of all one requires that an anti-ferromagnetic ground state: 
\begin{equation}    
 |M\rangle^{(\infty)} = u_M\wedge u_{M-1}\wedge u_{M-2} \wedge \cdots  \label{e: www}
\end{equation}
be a vector in the semi-infinite wedge product. Then the  rest of the vectors that span the product are obtained by perturbations of the sequence $ M, M-1,M-2,\dots $ in (\ref{e: www}) in finite number of places. The wedge product so defined is interpreted as the space of states of the conformal limit of the Calogero-Sutherland Model over the anti-ferromagnetic vacuum, and can be identified with a Fock space of two complex fermions.    

The generators of the algebras $\slt$ and $H$ remain well-defined operators in this space of states \cite{KMS}. So do generators of the Yangian and all the generators of the Virasoro algebra except $L_0$. After a minor redefinition necessitated by a subtraction of the infinite ground-state energy, the Calogero-Sutherland Hamiltonian and $L_0$  also become well-defined.

The fermion Fock space can be decomposed into  a tensor product of an irreducible level-1 representation of $\slt$ and the bosonic Fock space representation of the Heisenberg algebra $H$ \cite{KMS}.Therefore the Calogero-Sutherland Hamiltonian and the Yangian generators can be written in terms of $\slt$-generators and the bosons -- generators of $H$.
 At some special values of the coupling constant $\alpha$ the boson part of the fermion Fock space can be projected out. The corresponding projections of the Hamiltonian and the Yangian generators reproduce the Haldane-Shastry Hamiltonian and its Yangian symmetry algebra acting in one of the irreducible level-1 representations of $\slt$. This gives a connection with the previous, well-known  work of \cite{BPS,BLS}.       

It may be worthwhile to emphasize, that the results of this section are {\it derived} from those described in sec.1, all the intermediate steps being made transparent by means of using the wedge formalism.

\subsection{Semi-infinite wedge product}

Charge $M$ semi-infinite wedge product of spaces $V(z)$ , which will be denoted by $ F^{(M)} $ , is defined as a linear space with a basis formed by elements:\begin{eqnarray}
     u_{k_1}\wedge u_{k_2} \wedge \ldots   \quad, & &  \qquad (k_1 > k_2 > \ldots ), \\  k_i = M - i + 1 \quad  \text{ when } \quad  i \gg 1 \;.  \nonumber \label{e: siwedge} 
\end{eqnarray}
The vector $| M \rangle $ $\in $ $ F^{(M)} $ :
\begin{equation}
 | M \rangle  = u_M \wedge u_{M-1}\wedge u_{M-2} \wedge \ldots   \qquad \in F^{(M)} \,, \label{e: vacuum}
\end{equation}
will be called a {\it vacuum } vector of charge $M$.
Let $F$ be the  direct sum of spaces $F^{(M)}$ with all integer charges: 
\begin{equation}
 F = \bigoplus_{M \in \zint } F^{(M)}  \, . 
\end{equation}
The space $F$ can be identified with the fermion Fock space \cite{KR,KMS}. Introduce  wedging operators $\bar{\psi}_k$ : $ F^{(M)} \rightarrow F^{(M+1)}$, and contracting operators $ \psi_k $ : $ F^{(M)} \rightarrow F^{(M-1)}$, $ \in \zint $  by their action  on a semi-infinite wedge $w$ :
\begin{eqnarray}
 & & \bar{\psi}_k .w \;  = \; u_k \wedge w \quad , \label{e: fermions} \\ 
 & & \psi_k .( u_{k_1}\wedge u_{k_2}\wedge \ldots \;)   = \nonumber \\ & & = \begin{cases}
0 & \text{if $ k\not= k_i $  , $ i \geq 1 ,$ }\\  
(-1)^{k-1} ( u_{k_1}\wedge u_{k_2}\wedge \dots \wedge \widehat{u_{k_i}} \wedge \ldots \;) & \text{if $ k = k_i $ }. \end{cases} \nonumber  
\end{eqnarray}
where the hat over $u_{k_i}$ indicates that this factor is omitted from the product. 

The operators $\bar{\psi}_k$ and $ \psi_k$ generate the entire space $F$ from the vacuum $|0\rangle $ and satisfy the usual fermion anticommutation relations: 
\begin{eqnarray}
\{\bar{\psi}_k , \psi_l \} & = & \delta_{kl}\;, \label{e: anicomm}  \\
\{\bar{\psi}_k ,\bar{\psi}_l \} & = & \{\psi_k , \psi_l \} \; =\;0 \;. \nonumber
\end{eqnarray}

Sometimes it is covenient to consider the fermions with odd and even modes separately. Then $\bar{\psi}_{2k+\ep} \;( \psi_{2k+\ep}) $ are identified with creation operators of electrons ( holes ) with spin up ($ \ep = 1 $) or down ($\ep =2 $). 

\subsection{Action of the algebras $\slt$ , $H$ and $Vir$ in the semi-infinite
wedge product}

The space of semi-infinite wedges $F^{(M)}$ becomes an  $\slt$-module after one defines an action of the generators $\{J_m^a\}$ in $F^{(M)}$  by replacing the finite summation in the formula (\ref{e: affsl2N}) with an infinite one. Even though the $\slt$ generators are expressed by infinite sums, they map a semi-infinite wedge into a {\it finite } linear combination of semi-infinite wedges. 
Therefore these generators are well-defined operators in $F^{(M)}$ and as such can be written as normal-ordered  fermion expressions:   
\begin{eqnarray}    
J_m^a \quad & = & \quad \lui{J}_m^a \quad = \quad \sum_{i=1}^{\infty} z_i^m \s^a_i\quad = \label{e: affsl2i}\\ &= & \sum_{k\in\zint}\sum_{\ep_i=1,2} \s_{\ep_1,\ep_2}^a \bar{\psi}_{\ep_1-2k-2m}\psi_{\ep_2-2k} \; ,\nonumber   \\
 & & \text{where $ \s^a_{\ep',\ep}$ is defined by:} \quad       \s^a v_{\ep}\; = \;  v_{\ep'}\s^a_{\ep',\ep} \;. \nonumber 
\end{eqnarray}

Similarly, starting from (\ref{e: BBN})  one defines in $F^{(M)}$ an action of the Heisenberg algebra: 
\begin{eqnarray}
B_m & = & \lui{B}_m \quad = \quad \sum_{i=1}^{\infty} z_i^m \quad = \qquad (m \in \zint_{\not=0})\label{e: BBi}  \\ & &  = \; \sum_{k\in\zint} \bar{\psi}_{k-2m}\psi_k \; ,  \nonumber  \\
& & B_0 \quad = \quad \sum_{k\in\zint} :\bar{\psi}_{k}\psi_k: \; , \nonumber  \\
& & B_0 |_{F^{(M)}}  \quad = \quad M \mathrm{Id}|_{F^{(M)}} \; . \label{e: B0} 
\end{eqnarray}
The normal ordering $:\;:$ which appears above in the expression for the charge operator $B_0$ is taken with respect to the vacuum vector $| 0 \rangle $.

The transition from the finite to the semi-infinite wedge product has two important  effects. First, the commutation relations of $ \slt $ and $H$ acquire anomalous terms:  $F^{(M)}$ is a {\it level-1} representation of $ \slt $ , and the power sums $\{B_n\}$ , $ n\in \zint_{\not= 0} $ become creation ($ n < 0 $) and annihilation ($ n > 0 $ ) operators of bosons. The commutativity between $\slt$ and $H$ actions , however, remains intact:
\begin{eqnarray}
  & \quad [ J_m^a , J_n^b]& \; = \; 2i\ep^{abc}J_{m+n}^c \; + \; 2 m \delta^{ab}\delta_{m+n,0}  \;, \label{e: JJ} \\
  & \quad [ B_m , B_n]& \; = \;  2 m \delta_{m+n,0}  \;, \label{e: BB} \\
& \quad [ J^a_m , B_n]& \; = 0 \;  .  \label{e: JB}
\end{eqnarray}

The second effect of going over to semi-infinite wedges is that unlike the finite wegde product, $ F^{(M)}$ is a {\it highest-weight } representation of both $\slt $ and the Heisenberg algebra -- the vacuum $|M\rangle$ is the highest-weight vector: 
\begin{eqnarray}
J_n^a |M \rangle & = & B_n |M \rangle \quad = \quad 0 \;, \quad n > 0 \label{e: BJhw} \\ J_0^3 |M \rangle & = & \begin{cases} 0 & \text{ $ M $ is even ,} \\ |M\rangle & \text{ $ M $ is odd .} \end{cases} \label{e: J3M}
\end{eqnarray}

The space $F^{(M)}$ is reducible with respect to both $\slt $ and $H$ actions. However, it is an irreducible module of the direct sum $ \slt \oplus H $. The decomposition of $F^{(M)}$ into irreducible representations of the affine $\sll$ and the Heisenberg algebras is well known \cite{KMS}: 
Let $ V(\Lambda_i) $ , $ i=0,1$ be  the two level-1 irreducible, integrable $\slt$ representations with $\slt'$ highest weights $ \Lambda_i $ and highest-weight vectors $ |\Lambda_i\rangle $; and $ \cplx[H_{-} ]$ = $ \cplx[B_{-1},B_{-2},\dots] $ be the Fock space of the bosons $\{ B_n \}$. Then $ F^{(M)}$ is isomorphic as $ \slt \oplus H $ representation to the tensor product of  $ V(\Lambda_i) $ and $ \cplx[H_{-} ]$:
\begin{eqnarray}
     F^{(M)} & = & V(\Lambda_i) \otimes \cplx[H_{-} ] \; \label{e: decomp}  \\
     | M \rangle & = & |\Lambda_i\rangle \otimes 1 \; \label{e: vdecomp} \\
& & \text{ where $ i = 0(1) $ when $ M $ is even(odd) } \nonumber
\end{eqnarray}

Since $F^{(M)}$ is an irreducible representation of $ \slt \oplus H $, any operator acting in $F^{(M)}$ can be expressed in terms of the operators $\{ J_n^a , B_n \}$, $a=1,2,3 $, $ n \in \zint $.

Now let us turn to the Virasoro algebra which was defined in the finite wedge product by the expressions (\ref{e: VirN}). Replacement of the finite summation in (\ref{e: VirN}) by an infinite one, as one goes from the finite to the semi-infinite wedge product,  gives rise to well-defined operators $L_m$ when $m\not= 0$. 
In the case of $L_0$ , in order to obtain a well-defined operator in $F^{(M)}$, one has to subtract term by term vacuum $|0\rangle $ eigenvalues of the differentials $\{ D_i \}$:   
\begin{eqnarray}
L_m & = &  \lui{L}_m \quad = \quad -\sum_{i=1}^{\infty} z_i^m D_i\quad = \quad (m \in \zint_{\not=0})\label{e: Viri} \\ & &  = \; -\sum_{k\in\zint} k  \sum_{\ep=1,2}\bar{\psi}_{\ep-2k-2m}\psi_{\ep-2k} \; ,   \nonumber \\ & & 
L_0 \quad = \quad -\sum_{i=1}^{\infty} ( D_i - h^{(0)}_i) = \label{e: L0} \\ & & = \; -\sum_{k\in\zint} k  \sum_{\ep=1,2}\::\:\bar{\psi}_{\ep-2k}\psi_{\ep-2k}\: : \; .  \nonumber  
\end{eqnarray}
The numbers $ {h_i^{(0)}} $ that enter into the formula (\ref{e: L0}) are defined as follows: for any $ M $ put:
\begin{eqnarray}
& & \label{e: vacuumheights}\\
|M\rangle = u_M\wedge u_{M-1}\wedge u_{M-2} \wedge \ldots  & \equiv &  z^{h^{(M)}_1}v_{\ep_1}\wedge z^{h^{(M)}_2}v_{\ep_2}\wedge z^{h^{(M)}_3}v_{\ep_3}\wedge \ldots \; , \nonumber \\ 
               ( \ep_i\quad &  = & M + i  - 1 \quad ( \text{mod 2} )). \nonumber 
\end{eqnarray}

The action of $Vir$ in $F^{(M)}$ so defined has a central charge equal to $-4$ ; and is a highest-weight action, with the highest-weight vector $|M\rangle $:

\begin{eqnarray}
L_n |M\rangle &  = &  0 \; , \qquad n > 0 \; , \nonumber \\ 
L_0 | M\rangle &  = &  \begin{cases} \frac{M}{2}(\frac{M}{2}-1)| M\rangle
& \text{ $ M $ is even, } \\
(\frac{M -1 }{2})^2| M\rangle
& \text{ $ M $ is odd. } \end{cases} \label{e: L0M}
\end{eqnarray}

The Virasoro generators $\{L_m\}$, can be expressed in terms of $\{ J_n^a \}$ and $\{ B_n \}$ as guaranteed by  (\ref{e: decomp}). In order to obtain this expression one can use the commutators: 
\begin{equation}   
   [ L_m , J_n^a ] \quad = - n\;\; J_{n+m}^a \; ,  \label{e: LJi}
\end{equation}
and 
\begin{equation}  
[ L_m , B_n ] \quad = - n\;\; B_{n+m} + (1-m)m\delta_{m+n,0} \; . \label{e: LBi}
\end{equation}
Notice the presence of an anomalous term in (\ref{e: LBi}) as compared to (\ref{e: LBN}). 

These commutation relations together with the irreduciblity of $F^{(M)}$ as $\slt \oplus  H$ representation give rise to the following decomposition of $L_m$ , $ m \in \zint $ : 
\begin{equation}
\overset{c = -4}{L_m} \quad = \quad \overset{c = 1}{\LJ_m}\quad + \quad \overset{c =-5}{\LB_m}\;,\label{e: VirDec}
\end{equation}
where $\{\LJ_m \}$ denote the {\it Sugawara operators } constructed from the $\slt$ generators $\{ J_m^a \}$ ; and $\{\LB_m \}$ are the ``free boson`` generators of $Vir$ \cite{KR}:   
\begin{equation}
\LB_m = \frac{1}{4}\sum_{n\in\zint}\vdots B_{-n}B_{n+m}\vdots \; + \; \frac{1}{2}(m-1)B_m \,.\label{e: BVir}
\end{equation}
For any two $ A_m $ , $ C_n $ the conformal field theory normal ordering $ \vdots \; \vdots $ is defined as follows:
\begin{equation}
\vdots A_m C_n \vdots \; = \; \begin{cases} A_m C_n &  n \geq m \; , \\
 C_n A_m &  n < m \; . \end{cases} \label{e: CFTno}
\end{equation}

The Sugawara operators $\{\LJ_m \}$ have central charge 1 ; and the operators  
 $\{\LB_m \}$ have  central charge equal to $ -5$. The last operators belong to the ``Coulomb gas `` family of Virasoro representations which have central charges $ 1 - 6 \mu^2 $ , $ \mu \in \cplx $.   

The vacuum vector $|M \rangle $ = $ |\Lambda_i \rangle \otimes 1 $ is still a highest weight vector of both $\{\LJ_m \}$ and $\{\LB_m \}$:

\begin{eqnarray}
\LJ_n |M\rangle & = & \LB_n |M\rangle \quad = \quad 0 \;,\quad n > 0 \label{Virhw} \\ 
\LJ_0 |M\rangle & = & \begin{cases} 0 & \text{ $ M $ is even ,} \\ \frac{1}{4}|M\rangle & \text{ $ M $ is odd .} \end{cases} \label{e: LJ0M} \\
\LB_0 |M\rangle & = & \frac{M}{2}(\frac{M}{2}-1)| M\rangle \; .\label{e: LB0M} 
\end{eqnarray}

\subsection{Action of the Yangian and the fermion  spin-$\frac{1}{2}$  Calogero - Sutherland Hamiltonian in the semi-infinite wedge product $F$}
\mbox{}

Now let us turn to the Yangian generators that were defined in the finite wedge product by (\ref{e: Yangian0N},\ref{e: Yangian1N}) and the generalized Calogero-Sutherland Hamiltonian defined by (\ref{e: HCSN}). Our aim in this subsection is to transplant these objects from the finite to the semi-infinite wedge setting. 

Consider first the  Yangian. In the Yangian generators  we replace the finite summations in (\ref{e: Yangian0N},\ref{e: Yangian1N}) by infinite ones: 
\begin{eqnarray}
Q_0^a & \quad = \quad & \lui{Q}_0^a \; = \; J_0^a \; , \label{e: Yangian0i}\\
Q_1^a & \quad = \quad & \lui{Q}_1^a . \label{e: Yangian1i}
\end{eqnarray}



In the case of the generators $\lui{Q}_1^a$ one can be more specific and, replacing in (\ref{e: YNw}) $N$ with $\infty$,  write an action of these operators  on an {\it ordered } semi-infinite wedge $w$:   
\begin{equation}
w \; = \; z^{m_1}v_{\ep_1} \wedge z^{m_2}v_{\ep_2}\wedge z^{m_3}v_{\ep_3}\wedge\ldots\quad \in F^{(M)}  \label{e: genericinfwedge}
\end{equation}
This action is given by:
\begin{eqnarray}
& & \label{e: Yangian1iw} \\
& & \lui{Q}_{1}^a .w = \sum_{i=1}^{\infty}( \alpha m_i \s_i^a  -i \s^a_i) w + \nonumber\\ & & + \sum_{i=1}^{\infty}\delta_{m_i,m_{i+1}}(\s_i^a - \s_{i+1}^a). w + \frac{i}{2} \sum_{1\leq i < j \leq \infty}\ep^{abc} \s_i^b \s_j^c . w  -  
\sum_{1\leq i < j \leq \infty} q^a_{ij}. w  \nonumber 
\end{eqnarray}
Despite the presence of double infinite sums in the last expression, it specifies  well-defined operators in a space of semi-infinite wedges $F^{(M)}$ -- that is acting with $ \lui{Q}_{1}^a $ on a wedge  produces a finite linear combination of wedges.

Similarly, an action of the generalized Calogero-Sutherland Hamiltonian in $F^{(M)}$ is obtained by extending the sums in (\ref{e: HNw}) to all positive integers. In this case, however, in the single sums in  (\ref{e: HNw}) one has to subtract term by term vacuum eigenvalues of the summands -- as it had been done for $L_0$ in the previous subsection. Thus one defines an operator  $H_{CS}(\alpha)$ :
\begin{equation}
H_{CS}(\alpha) \quad = \quad \lui{H}_{CS}(\alpha) \;, \label{e: HCSi}
\end{equation}
where the action of $\lui{H}_{CS}(\alpha)$ on an {\it ordered } wedge $w$ (\ref{e: genericinfwedge}) is: 
\begin{equation}
\lui{H}_{CS}(\alpha).w = \sum_{i=1}^{\infty}\{ \alpha (m_i^2 -(h_i^{(0)})^2) + 2i (m_i - h_i^{(M)}) \}w + 2\sum_{1\leq i < j \leq \infty} h_{ij}. w  \; . \label{e: HCSiw}
\end{equation}
Like the Yangian generators $\{\lui{Q}_{0}^a , \lui{Q}_{1}^a \} $, the operator  $\lui{H}_{CS}(\alpha)$ is well-defined in $ F^{(M)} $.

Thus we have defined an action of the Yangian $Y(\slt)$ and an action of the fermionic, spin-$\frac{1}{2}$ Calogero-Sutherland Hamiltonian in each of the spaces $F^{(M)}$ and, consequently, in the Fock space $F$. 
Strictly speaking one must check, that the operators $\{\lui{Q}_{0}^a , \lui{Q}_{1}^a \} $ still satisfy the Yangian defining relations and commute with $\lui{H}_{CS}(\alpha)$ -- as  going over to semi-infinite wedges may give rise to  anomalies in the commutation relations. In the next section we shall see, however, that such ``symmetry breaking'' anomalies do not appear. 

As the next step, one can express the Yangian generators and the Hamiltonian in terms of the $\slt$ generators $\{ J_n^a\}$ and the bosons $\{ B_n \}$ in accordance with the decomposition (\ref{e: decomp}). For the Yangian one obtains the following expressions, where the notation is chosen so as to indicate the dependence of the Yangian generators on the parameter $\alpha$: 
\begin{eqnarray}
& & \label{e: Y0} \\
Q_0^a(\alpha) &  = & \lui{Q}_0^a\; = \; J_0^a \quad , \nonumber \\
& & \label{e: Y1} \\
Q_1^a(\alpha)  &  = & \lui{Q}_1^a\; = \; (1+\frac{\alpha}{2}(1-M)) J^a_0 -  \nonumber \\  & & - \frac{\alpha}{2}\sum_{n \geq 1} J_{-n}^a B_n - (\frac{\alpha}{2}+1)\sum_{n \geq 1} B_{-n} J_n^a  -\frac{i}{2}\sum_{n \geq 1}\ep^{abc} J_{-n}^b J_{n}^c \quad. \nonumber
\end{eqnarray}

For the Hamiltonian one obtains:
\begin{eqnarray}
& & \label{e: Hamiltonian} \\
H_{CS}(\alpha) & = & \quad  \lui{H}_{CS}(\alpha) \quad  =  \nonumber \\  
& & =\quad \frac{1}{12}M(2 M^2 - 3 M - 2)\quad +\quad \frac{\ep}{4}\quad -\quad M L_0 + \nonumber \\  
& &  + \quad (\alpha + 1)(M - 1)(\frac{1}{3}\LB_0 \quad +\quad \LJ_0 ) \quad+ \nonumber \\ 
& & +\quad \frac{(\alpha + 4)}{3} \sum_{n \geq 1} B_{-n}\LB_n\quad +\quad \frac{(\alpha -2)}{3} \sum_{n \geq 1}\LB_{-n}B_n \quad+ \nonumber \\  
& & +\quad (\alpha + 2 ) \sum_{n \geq 1} B_{-n} \LJ_n\quad + \quad\alpha \sum_{n \geq 1} \LJ_{-n}B_n\quad - \nonumber \\
& &  -\quad\frac{3}{2}\sum_{n \geq 1} n  B_{-n}B_n\quad - \quad \frac{1}{2}\sum_{n \geq 1}  \sum_{a=1}^{3} n J_{-n}^a J_{n}^a  \; ,\quad \nonumber \\
& &  \text{where $ \ep $ = 0(1) when $M$ is even(odd). } \nonumber 
\end{eqnarray}

It is clear that $L_0 = \LJ_0 + \LB_0$ commutes both with the Yangian and with the Hamiltonian. So, to simplify $H_{CS}(\alpha)$ we shall, in what follows, remove the constant term and the term proportional to $L_0$, defining $H_{CS}(\alpha)'$ by: 
\begin{equation*}
H_{CS}(\alpha)  = H_{CS}(\alpha)' + \frac{1}{12}M(2 M^2 - 3 M - 2) + \frac{\ep}{4} - M L_0 \; ,
\end{equation*}
and from now on dropping the prime. 

The equations (\ref{e: Y0},\ref{e: Y1}) and (\ref{e: Hamiltonian}) constitute the main result of this section. They define the conformal limit of the fermionic spin-$\frac{1}{2}$ generalized Calogero-Sutherland Hamiltonian and the generators of the associated Yangian symmetry.    

It is natural to call the $\slt$ generators ``spin'' degrees of freedom and the bosons ``charge'' degrees of freedom. The decomposition (\ref{e: decomp}) then can be interpreted as a  ``spin-charge'' separation of the space of states into a product of pure spin and pure charge factors \cite{BPS}. This separation  is a feature of the conformal limit and is absent in the situation where the number of particles in the Calogero-Sutherland Model is finite. 

 At generic values of the parameter $\alpha $ the space $F^{(M)}$ is the space of states of the model and cannot be reduced to smaller spaces without breaking the property of integrability and the Yangian invariance. At two  special values of the parameter $\alpha $, however, such a reduction is possible. These values are: 
\begin{equation}
 \alpha = 0 \; , \; \text{and} \; \;  \alpha = -2 \; . \label{e: SpecialPoints}
\end{equation}

First consider the point $ \alpha = 0 $. Let $ F^{(M)'} $ be a subspace of $F^{(M)}$ which consists of all $ V(\Lambda_i)$-valued  polynomials without constant term in the boson creation operators, i.e:
\begin{equation}
F^{(M)'} \quad = \quad (\oplus_{n \geq 1} \cplx B_{-n}) F^{(M)} \; .
\end{equation}
We have an isomorphism of $\slt$ modules:
\begin{equation}
F^{(M)}/ F^{(M)'} \simeq  V(\Lambda_{M \bmod 2})\otimes 1 \simeq  V(\Lambda_i)\; .  \label{e: quotient}
\end{equation}
The operators $Q_1^a(0) $ and $ H_{CS}(0) $ preserve the space $ F^{(M)'} $:
\begin{equation}
Q_1^a(0)\;,\; H_{CS}(0) \; : \quad F^{(M)'} \hookleftarrow \quad . \label{e: pre1}
\end{equation}
Indeed, $Q_1^a(0)$ depends only on the creation operators $\{B_{-n}\}$, $ n > 0 $, whereas $ H_{CS}(0) $ contains the annihilation operators $\{B_{n}\}$, $ n > 0 $ only in combinations:
\begin{equation*}
B_{-n}B_n \; , \; B_{-n}\LB_n \; , \; \LB_{-n}B_n \; ,
\end{equation*}
all of which commute with $\LB_0$ and therefore preserve the principal degree of polynomials in $\{B_{-n}\}$, $ n > 0 $. 

Thanks to (\ref{e: pre1})one  can quotient out the subspace  $ F^{(M)'} $ from the space of states and restrict the model upon $ V(\Lambda_i)$ as can be seen from (\ref{e: quotient}). A convenient way to implement this quotient is to put all the creation operators of bosons equal to zero: 
\begin{equation}
 B_n   =   0 \;, \quad   n < 0 \;. \label{e: B-=0} 
\end{equation}
Let $\iota$ be the map :
\begin{equation}
\iota \;:\; End(F^{(M)}) \rightarrow  End(V(\Lambda_{M\bmod 2})) \;  
\end{equation}
induced by the quotient map (\ref{e: quotient}). Then the operators that define the restriction of spin Calogero-Sutherland Model upon the highest-weight irreducible $\slt$ representations $ V(\Lambda_i)$ are seen to be:
\begin{eqnarray}
& & \label{e: YHS}\\
\iota(Q_0(0)) & = & J_0^a \; ,\nonumber \\
\iota( Q_1^a(0) ) &  = &  J_0^a  - \frac{i}{2}\sum_{n\geq 1}\ep^{abc} J_{-n}^b J_{n}^c  \; ,\nonumber  \\  & & \label{e: HS} \\ 
\iota( H_{CS}(0) ) & = & \frac{1}{12}M(M-1)(M - 2) +  (M-1) \LJ_0 -  \frac{1}{2}\sum_{n \geq 1}  \sum_{a=1}^{3} n J_{-n}^a J_{n}^a  \; . \nonumber 
\end{eqnarray}
In these operators one  recognizes the Yangian generators and the Haldane-Shastry Hamiltonian which were discovered in \cite{HHTBP} and further studied in \cite{BPS} and \cite{BLS}.

Another point where one can reduce the space of states $F^{(M)}$ is the point $ \alpha = -2 $. In this case the subspace $ V(\Lambda_{M\bmod 2})\otimes 1 $ is invariant under the action of $Q_1^a(-2)$ and $ H_{CS}(-2)$:
\begin{equation}
Q_1^a(-2)\;,\; H_{CS}(-2) \; : \quad V(\Lambda_{M\bmod 2})\otimes 1 \hookleftarrow \quad . 
\end{equation}

The restrictions of $Q_1^a(-2)$ and $ H_{CS}(-2)$ upon $V(\Lambda_{M\bmod 2})\otimes 1$ again reproduce the Haldane-Shastry Hamiltonian and  the associated Yangian generators: 
\begin{eqnarray}
& & \\
 Q_1^a(-2) |_{V(\Lambda_{M\bmod 2})} &  = & M J_0^a  - \frac{i}{2}\sum_{n\geq 1}\ep^{abc} J_{-n}^b J_{n}^c  \; ,\nonumber  \\  & &  \\ 
 H_{CS}(-2)|_{V(\Lambda_{M\bmod 2})}  & = & -\frac{1}{12}M(M-1)(M - 2) -  (M-1) \LJ_0 - \nonumber \\ & & - \frac{1}{2}\sum_{n \geq 1}  \sum_{a=1}^{3} n J_{-n}^a J_{n}^a  \; . \nonumber 
\end{eqnarray}
At this point the constraints that lead to the reduction of the space of states are  opposite  to (\ref{e: B-=0}) -- they are expressed by putting the annihilation operators of bosons equal to zero:
\begin{equation}
 B_n   =   0\;,  \quad   n > 0 \;. \label{e: B+=0} 
\end{equation}

One can trace the origin  of the constraints (\ref{e: B-=0},\ref{e: B+=0}) to the situation of finite $N$.
To do this let us recall, following \cite{TH}, that in the case of finite number of particles the Haldane-Shastry Model can be obtained from the fermion Calogero-Sutherland Model (\ref{e: HCSFN}) at $\alpha = 0$ by pinning down the coordinates of the particles $\{ z_i \}$ in an equally-spaced fashion around the unit circle in the complex plane:
\begin{equation}
   z_j = \exp( \frac{2\pi ij}{N} ) \; , \quad j = 1,\dots, N \;. \label{e: constr1}
\end{equation}
To be more precise, the above constraints must be amended because a multiplication operator $ z_j $ is not defined in the space of states of fermions -- its action violates the antisymmetry condition:
\begin{equation}
 K_{ii+1} = - P_{ii+1} .\; 
\end{equation}
Rather, instead of (\ref{e: constr1}) one may substitute the weaker requirement that the power sums (\ref{e: BBN}) vanish:    
\begin{equation}
 \luN{B}_m  = \sum_{i=1}^{N} z_i^m = 0 \;, \quad m \not= 0 (\bmod N). \label{e: constr2}
\end{equation}
Since the power sums transform into bosons in the semi-infinite wedge limit (\ref{e: BBi}), by analogy with (\ref{e: constr2}) the constraints (\ref{e: B-=0},\ref{e: B+=0}) can be now taken to  express a  pinning of an ``infinite number `` of particles along the unit circle, so that only their spin degrees of freedom survive. 

Finally, let us remark that at the special points $ \alpha = 0 \;, \; -2 $ one can eliminate the spin degrees of freedom so that the Yangian generators become equal to zero and in the Hamiltonian only the terms which depend exclusively on the bosons remain. The Hamilonian so obtained resembles, but is not identical to the spinless Calogero-Sutherland Hamiltonian acting in the Fock space of bosons which was recently defined and studied in \cite{Awata}.

\section{Decomposition of $F^{(M)}$ into irreducible representations of the Yangian}

In this section we shall demonstrate, that at generic values of the parameter $\alpha $ the space of semi-infinite wedges $F$ or, equivalently, the Fock space of fermions  (\ref{e: fermions}) is decomposed into a direct sum of irreducible representations of the Yangian defined by the generators (\ref{e: Y0},\ref{e: Y1}). 
These representations are obviously eigenspaces of the Hamiltonian (\ref{e: Hamiltonian}). Since the Yangian generators and the Hamiltonian act within  each of the subspaces of fixed charge $F^{(M)}$, the problem is reduced to a decomposition of $F^{(M)}$.   

\subsection{ A ``fermion'' basis of $F^{(M)}$ }
 
As a starting point in the decomposition we define a suitable basis in $F^{(M)}$. This basis is essentially the basis formed by the ordered wedges represented in a convenient form. We call this basis a ``fermion'' basis by analogy with the spinon basis of ref. \cite{BPS,BLS}.

 Let $\Om \; \subset \;F^{(M)}$ be a subspace of $F^{(M)}$ defined with the aid of the creation operators of holes $\psi_k$ as follows:  
\begin{eqnarray}
 &  & \label{e: Om} \\  
\Om & = & \cplx\{\psi_{\ep_1 -2 (a_1 + s)}\psi_{\ep_2 -2 (a_2 + s)}\dots \psi_{\ep_N -2 (a_N + s)}|N + M\rangle \}\begin{Sb} a_i \in \zint \\ \ep_i = 1,2 \end{Sb} ,  \nonumber \\ & &  s \; = \; 1 - \frac{N + M}{2} \; . \nonumber 
\end{eqnarray}

The spaces $\Om$ with $N$ such that $N+M$ is an even number span the entire space $F^{(M)}$: 
\begin{equation}
F^{(M)} = \bigcup\begin{Sb} N \geq 0 \\ N+M = \text{even} \end{Sb} \Om \; , 
\end{equation}
and from now on we shall consider $ \Om $ with $ N+M = \text{even} $ only.

The spaces $\Om$ with different values of $N$ are not, however, linearly independent. Rather, they are imbedded into each other as expressed by the following inclusion: 
\begin{equation}
   \Om \; \supset \Omm \; ,\quad N \geq 2 . \label{e: Inclusion} 
\end{equation}
Therefore:
\begin{eqnarray}
F^{(M)} & = & \bigoplus\begin{Sb} N \geq 0 \\ N+M = \text{even} \end{Sb} \Om\setminus \Omm \;, \label{e: dec1} \\ & & \text{where for $ N < 0 $ we put } \; \Om = 0 \;. \label{e: negOm} 
\end{eqnarray}

Let $w = (w_1,\dots,w_N) $ be formal variables and $ \{ v_1 , v_2 \}$ be the  basis in $ V = \cplx^2 $, $w$ and  $ \{ v_{\ep}\}$ will be called auxiliary variables.
Define a generating function $\Omw$ for the vectors which form the space $\Om$:
\begin{eqnarray}
& & \label{e: Omw} \\
\lefteqn{\Omw = }\nonumber \\
 &  & = \sum\begin{Sb} a_i \in \zint \\ \ep_i = 1,2 \end{Sb}\psi_{\ep_1 -2 (a_1 + s)}\dots \psi_{\ep_N -2 (a_N + s)}|N + M\rangle (v_{\ep_1}\otimes\dots\otimes v_{\ep_N}) w_1^{a_1}\dots w_N^{a_N} ,  \nonumber \\ & &  \Omw \; \subset \; \Om\bigotimes (\cplx[w_1,\dots,w_N] \otimes (\otimes^N V))_F  \; .\nonumber 
\end{eqnarray}
This generating function contains only non-negative powers of the variables $w_1, \dots , w_N $. This follows from:  
\begin{equation*}
\psi_{\ep -2 (a + s)}| N + M > = 0 \;, \quad a < 0 \;,\; \ep = 1,2 \; .
\end{equation*}

Moreover $\Omw$ is totally antisymmetric in the auxiliary variables:
\begin{eqnarray}
(\cplx[w_1,\dots,w_N] \otimes (\otimes^N V))_F &  \subset & \cplx[w_1,\dots,w_N] \otimes (\otimes^N V) \;, \nonumber \\ 
(\cplx[w_1,\dots,w_N] \otimes (\otimes^N V))_F & = & \bigcap_{1\leq i \leq N-1}Ker(K_{ii+1} + P_{ii+1})\; . \label{e: fermionspace}
\end{eqnarray}
In terms of the generating functions the inclusion  (\ref{e: Inclusion}) can be expressed as follows :
\begin{equation}
\Omw |_{w_{N-1}=w_{N}=0} = (w_1 w_2 \dots w_{N-2})\Omega_M^{(N-2)}(w)\otimes(v_1\otimes v_2 - v_2\otimes v_1) \;. \label{e: OmwInclusion}
\end{equation}
Computation of the action of the Yangian generators (\ref{e: Y0}, \ref{e: Y1}) on a generating functon $\Omw$ yields the following  result:  

\begin{eqnarray}
& &  \\
\lefteqn{Q_0^a(\alpha) \Omw  = } & & \hspace{2cm} - \{\sum_{i=1}^N \s_i^a \} \Omw \;, \label{e: Q0Omw} \nonumber  \\
& &  \\
\lefteqn{Q_1^a (\alpha)\Omw  = }  \nonumber   \\ & = & \{ \luN{\hat{Q}}_1^{a,F}(-(\alpha + 2)) + (\alpha(\frac{N+M}{2}-1)+N-1)\sum_{i=1}^N \s_i^a\}\Omw \; . \nonumber  \label{e: Q1Omw} 
\end{eqnarray}
In the formula for $Q_1^a(\alpha)\Omw $ we have used the notation (\ref{e: Yangian1NF}):
\begin{equation}
\luN{\hat{Q}}_1^{a,F}(\beta) = \sum_{i=1}^N d_i(\beta)\s_i^a  - \frac{i}{2} \sum_{1\leq i < j \leq N}\ep^{abc}\s_i^b \s_j^c \; . \label{e: fermionQ1}
\end{equation}
In the relations (\ref{e: Q0Omw},\ref{e: Q1Omw}) it is understood, that the operators $\{Q_0^a(\alpha), Q_1^a(\alpha)\}$ act in the first factor in the tensor product 
\begin{equation*}
\Om \bigotimes \FS \; ,
\end{equation*}
while the opertors which appear in the right hand side act in the second factor: $\FS$.  

Thus we see, that the action of the Yangian in $\Om$ is equivalently described by the action of the fermionic Yangian generators on the dual variables. However the coupling constant is different from the ``bare'' coupling constant $\alpha$ . The effect of working with semi-infinite wedges is a renormalization:      
\begin{equation}
\alpha \rightarrow \alpha + 2 \; .
\end{equation}
Also, there is a change of sign since we use the hole creation operators $ \psi_k $ to form the generating function $ \Omw$. From now on we shall fix the notation:
\begin{equation}
\beta = -( \alpha + 2 )\; .
\end{equation}

Similarly, the action of the Hamiltonian (\ref{e: Hamiltonian}) in $ \Om$ is reduced to the action of $N$-particle generalized Calogero-Sutherland Hamiltonian on the dual variables:

\begin{eqnarray}
& & \label{e: HOmw} \\
\lefteqn{ H_{CS}(\alpha) \Omw = } &  &  \nonumber  \\
& & = \{ \luN{H}_{CS}(\beta) + ((\beta +1)(2-N-M)-N-2)\sum_{i=1}^N D_i - \nonumber \\ 
& & \hspace{1.5cm} - \frac{(\beta +1)}{12}(N+M-2)(N+M-1)(N+M) + \nonumber \\ & & \hspace{3cm} + (\beta+1)N(1-\frac{N+M}{2})^2 \} \Omw  \; = \nonumber \\
& & =\{\frac{1}{\beta}\sum_{i=1}^N d_i(\beta)^2 + ((\beta +1)(2-N-M)+N)\sum_{i=1}^N D_i + \nonumber \\ & & \hspace{3cm} + c(\beta,N,M)\}\Omw\;; \nonumber \\ \lefteqn{ \text{where we put: }}\nonumber  \\ \lefteqn{ c(\beta,N,M) =}\nonumber \\ & & = -\frac{(\beta +1)}{12}(N+M-2)(N+M-1)(N+M) + (\beta+1)N(1-\frac{N+M}{2})^2 - \nonumber \\ & & \hspace{3cm} -\frac{1}{6\beta}N(N+1)(2N+1) \;,   \nonumber \\ \lefteqn{ \text{and $ \beta = -(\alpha  + 2) $}.}\; & &  \nonumber   
\end{eqnarray}

Three remarks are in order. First: the equations ( \ref{e: Q0Omw},\ref{e: Q1Omw})
and (\ref{e: HOmw}) demonstrate that the operators ( \ref{e: Y0},\ref{e: Y1}) satisfy the defining relations of the Yangian and commute with the Hamiltonian (\ref{e: Hamiltonian}).
 Second : the Dunkl opertors $\{ d_i(\beta) \}$ preserve the space of polynomials in $ w $, and therefore the equations ( \ref{e: Q0Omw},\ref{e: Q1Omw},\ref{e: HOmw}) contain only non-negative degrees of $w$ in both sides. Third: a  straightforward verification shows, that the equations (\ref{e: HOmw}) and (\ref{e: Q0Omw},\ref{e: Q1Omw}) are consistent with the restriction (\ref{e: OmwInclusion}).    

Finally, the action of the operator $L_0$ on $\Omw$ is found to be:
\begin{equation}
L_0\Omw = \{ \sum_{i=1}^N D_i + (\frac{M+N}{2}-1)\frac{(M-N)}{2} \}\Omw \; . \label{e: L0Omw}
\end{equation}

\subsection{Irreducilble decomposition of the space $\Om$ with respect to 
the Yangian action} 

The relations  ( \ref{e: Q0Omw},\ref{e: Q1Omw}) show, that the Yangian decomposition of the space $\Om$ will be found once we decompose the space $\FS$ into irreducible representations of the Yangian generated by the operators:
\begin{eqnarray}  
 & &  \\
\luN{\check{Q}}_{0,F}^a(\beta ) & = & - \sum_{i=1}^N \s_i^a  \; ,\nonumber  \label{e: Q0} \\  & & \\ 
\luN{\check{Q}}_{1,F}^a(\beta ) & = & \luN{Q}_1^{a,F}(\beta) + ((\beta +  2)(-\frac{N+M}{2}+1)+N-1)\sum_{i=1}^N \s_i^a \; . \nonumber   \label{e: Q1}  
\end{eqnarray}
\mbox{}From now on we assume, that $ \beta $ is a generic complex number, specifically not a rational.

At this point it will be convenient to switch to the $L$-operator formalism. Now we use another presentation of the Yangian generated by the operators (\ref{e: Y0}, \ref{e: Y1})-- the presentation in terms of the Yangian - valued matrix \cite{Drinfel'd}:
\begin{eqnarray}
 {\Bbb T}_0(x)  & \equiv & {\Bbb T}_{ij}(x) , \; i,j=1,2 ;\quad  \text{ det}_q {\Bbb T}(x) = 1 \;, \label{e: TT} \\
{\Bbb T}_0(x) & \in & End( V_0 )\otimes Y(\sll) \; , \; V_0 = \cplx^2 . \nonumber 
\end{eqnarray}

Introduce an $L$-operator:
\begin{equation}
L_{0i}(x) = I - \frac{P_{0i}}{x} \; ,\label{e: Loperator} 
\end{equation}
where $P_{0i}$ is the permutation operator of spaces $ V_0 $ and $ V_i $,  and $x$ is a spectral parameter.

For commuting operators or $c$-numbers $\{b_i\}$, $ \iN $ define {\it T-matrix } by: 
\begin{equation}
T_0(x,\{b_i\}) = L_{01}(x-b_1)L_{02}(x-b_2)\dots L_{0N}(x-b_N)\; . 
\end{equation}

Now the relations ( \ref{e: Q0Omw}, \ref{e: Q1Omw}) assume the following form:

\begin{eqnarray}
 & & \label{e: Tmatrix} \\
{\Bbb T}_0(x)\Omw & = &  \varphi(x) T_0(x,\{d_i(\beta) + \eta(\beta,N,M)\})\Omw\; , \nonumber \\ 
& & \text{where}\; \eta(\beta,N,M) = N-1+(\beta+2)(1-\frac{N+M}{2}) \; . \nonumber 
\end{eqnarray}
The $  \varphi(x) $ above is a scalar, symmetric function of Dunkl operators. It is defined by the requirement that  the quantum determinant of the $T$-matrix : 
\begin{equation} 
T_0(x) = \varphi(x) T_0(x,\{d_i(\beta) + \eta(\beta,N,M)\}) \; ,\label{e: T0}
\end{equation}
 be equal to 1. The explicit form of   $  \varphi(x) $ is not needed in what follows because Yangian highest-weight vectors and associated Drinfel'd Polynomials do not depend on   $  \varphi(x) $.
 
The rest of this subsection is concerned primarily with the description of the decomposition of the space $\FS$ into irreducible subrepresentations of the $T$-matrix (\ref{e: T0}). The details of the derivation of the $q$-deformed, bosonic analog of this decomposition can be found in \cite{U}, the adaptation to the present $q \rightarrow 1$, fermionic situation is straightforward. The results concerning the Yangian decomposition of the space $\Om$ appear at the end of this subsection in (\ref{e: Omwexpansion} -- \ref{e: Omexpansion}). 

Let $\l$ be a partition of  length $ \leq N $ :
\begin{equation*}
\l = ( \l_1 \geq \l_2 \geq \dots \geq \l_N \geq 0 ) \; ,
\end{equation*}
and let $\Lambda_N$ be the set of all such partitions.

Writing a partition $\l$ as:
\begin{multline}
\l = ( \l_1 =\dots = \l_{m_1} > \l_{1+m_1}=\dots= \l_{m_2} > \ldots \label{e: m-s} \\
\ldots  > \l_{1+m_J} = \dots = \l_N \geq 0) \nonumber ;
\end{multline}
define $S^{\l}$ as a subset of permutations of $\{1,\dots,N\}$ such, that in the elements of $S^{\l}$ the original order of the sequence $1,2,\dots,N$ is  preserved within each of the following $ J + 1 $ subsets:
\begin{gather*}
\{1,2,\dots,m_1\} \\
\{1+m_1,2+m_1,\dots,m_2\} \\
\vdots \\ 
\{1+m_J,2+m_J,\dots,N\} \; , \\
( 0 = m_0 < m_1 < m_2 < \dots < m_J < N ) \;.
\end{gather*}
Elements of the  set $S^{\l}$ are  in one-to-one correspondence with distinct rearrangements of 
the partition $\l$.

Associate with a partition $\l$ a sequence of $N$ numbers $ \{ \zeta^{\l}_i \}$, $ \iN $:
\begin{equation}
\zeta^{\l}_i = \beta \l_i - i \; . \label{e: zeta}
\end{equation}

Next, define a polynomial $\Phi^{\l}(w) $ by:
\begin{equation}
\Phi^{\l}(w) = \left\{\prod_{i=0}^{J-1}\prod_{j=1+m_i}^{m_{i+1}}\prod_{k=i+1}^J (d_j(\beta)-\zeta^{\l}_{1+m_k})\right\}J_{\l,N}^{(\beta)}(w) \; ; \label{e: Phi}
\end{equation}
where $J_{\l,N}^{(\beta)}(w)$ is the Jack Polynomial of $N$ variables associated with the partition $\l$ \cite{M}.

The polynomial $\Phi^{\l}(w) $ is a common eigenfunction of the Dunkl operators:
\begin{equation}
d_i(\beta).\Phi^{\l}(w) = \zeta^{\l}_i\Phi^{\l}(w) \; . \label{e: Eigen}
\end{equation}

Let $\Lambda^{(2)}_N $ $ \subset $ $\Lambda_N $ be a set of all partitions that contain {\it no more } than two equal parts of any given value $\geq 0 $.

For $\l \in \Lambda^{(2)}_N $, let $ I^{\l} \subset \setN $ be the maximal subset of $\setN$ such, that:
\begin{equation}
 \l_i = \l_{i+1} \quad \text{for all $i \; \in \; I^{\l} $ .}
\end{equation}

With $\l \in \Lambda^{(2)}_N $ associate a subspace $W^{\l}$ of the tensor product $\otimes^N V$ by antisymmetrizing factors $i$ and $i+1$ for all $ i \in I^{\l} $:  
\begin{equation}
W^{\l} = \prod_{i\in I^{\l}} (1 - P_{ii+1})\otimes^N V \; . \label{e: W}
\end{equation}

Fix a $\l \in \Lambda_N $. For all $\s \in S^{\l}$ define an operator $\BY^{\l}(\s;w)$ $: \otimes^N V $ $ \rightarrow $ $ \cplx[w_1,\dots,w_N]\otimes ( \otimes^N V ) $ by the following recursion relation:
\begin{eqnarray} 
\BY^{\l}(\text{id};w) & = & \Phi^{\l}(w) \; , \label{e: Y}\\
\BY^{\l}((i,i+1)\s;w) & = & Y_{i,i+1}(\zeta_{\s_i}^{\l} -\zeta_{\s_{i+1}}^{\l})\BY^{\l}(\s;w) \; ;\quad \s , (i,i+1)\s \in S^{\l} \; ;\nonumber \\ 
\lefteqn{\text{here the  operator $ Y_{i,i+1}(u)$ is:} } & &  \nonumber \\
Y_{i,i+1}(u) & = & - \frac{(1 - u K_{ii+1})(1 + u P_{ii+1})}{1 - u^2} \; . \nonumber 
\end{eqnarray}
A remark on this definition must be made. The set $S^{\l}$ is {\it connected } in the following sense: for any element $\s$ $\in$ $ S^{\l}$ one can find a chain of pairwise permutations $ (i_r,i_r+1), (i_{r-1},i_{r-1}+1)$ $ \dots $ $ (i_1,i_1+1) $ such that: 
\begin{equation}
\s = (i_r,i_r+1).(i_{r-1},i_{r-1}+1). \cdots .(i_1,i_1+1)\{1,2,\dots ,N\}\;, \label{e: c} 
\end{equation}
and for any $  1 \leq k \leq r $ the permutation 
\begin{equation}
(i_k,i_k+1).(i_{k-1},i_{k-1}+1). \cdots .(i_1,i_1+1)\{1,2,\dots ,N\}
\end{equation}
belongs to the set $S^{\l}$. The chain (\ref{e: c}) in general is not unique, but the operator $\BY^{\l}(\s;w)$ is defined unambiguously by the recurrence relation (\ref{e: Y}) because $ Y_{i,i+1}(u)$ satisfy the Yang-Baxter equation and the unitarity condition: $ Y_{i,i+1}(u)Y_{i,i+1}(-u)=I$.

Next, for any $\l \in \Lambda^{(2)}_N $ define the following operator: 
\begin{equation}
\BU^{\l}(w) = \sum_{\s  \in S^{\l}} \BY^{\l}(\s;w) \; . \label{e: U}
\end{equation}
The operator $\BU^{\l}(w)$ maps $ W^{\l} $ into $ \FS $, and is a Yangian intertwiner:
\begin{equation}
T_0(x,\{d_i(\beta) + \eta(\beta,N,M)\}).\BU^{\l}(w) = \BU^{\l}(w)T_0(x,\{\zeta^{\l}_i + \eta(\beta,N,M)\}) \; . \label{e: Intertwiner}
\end{equation}

Next, consider the unique  vector of lowest spin in the space $W^{\l}$: 
\begin{multline}
v^{\l} = v_2\otimes v_2\otimes \dots \label{e: vm} \\ \dots \otimes v_2\otimes(\underset{i_1}{v_1}\otimes \underset{i_1+1}{v_2} - \underset{i_1}{v_2}\otimes \underset{i_1+1}{v_1})\otimes v_2\otimes v_2\otimes \dots \nonumber \\ \dots \otimes v_2\otimes(\underset{i_2}{v_1}\otimes \underset{i_2+1}{v_2} - \underset{i_2}{v_2}\otimes \underset{i_2+1}{v_1})\otimes \ldots \nonumber \\ 
\ldots  \otimes v_2\otimes v_2 \; , \nonumber \\ 
i_1,i_2,\dots \in I^{\l} \; . \nonumber 
\end{multline}
Inspection of the numbers $\{ \zeta_i \}$ introduced in (\ref{e: zeta}) shows, that the matrix $ T_0(x,\{\zeta^{\l}_i + \eta(\beta,N,M)\})$ acts in the space 
$W^{\l}$ irreducibly, and $ v^{\l}$ is a lowest-weight vector of this matrix.

Now, fix a basis $\tau(\l)$ in $W^{\l}$. For $\tau \in \tau(\l) $ put: 
\begin{equation}
\phi_{\tau}^{\l}(w) = \BU^{\l}(w).\tau 
\end{equation}
The linear space $ F^{\l} $: 
\begin{equation}
F^{\l} = \oplus_{\tau \in \tau(\l)} \cplx\phi_{\tau}^{\l}(w) \; ,
\end{equation}
is an irreducible {\it lowest weight } representation of the Yangian generated by the $T$-matrix (\ref{e: T0}) with the lowest-weight vector $\phi_{v^{\l}}^{\l}(w)$:
\begin{eqnarray}
T_{12}(x).\phi_{v^{\l}}^{\l}(w) & = & 0 \; ,  \label{e: lw} \\
T_{11}(x).\phi_{v^{\l}}^{\l}(w) & = & t_{11}(x)\phi_{v^{\l}}^{\l}(w)\; , \nonumber \\
T_{22}(x).\phi_{v^{\l}}^{\l}(w) & = & t_{22}(x)\phi_{v^{\l}}^{\l}(w) \; . \nonumber
\end{eqnarray}
The Drinfel'd Polynomial $ P_N^{\l}(x) $ of this representation is defined by the ratio of the $c$-valued
functions $t_{11}(x),t_{22}(x)$:
\begin{equation}
\frac{t_{11}(x)}{t_{22}(x)} = \frac{P_N^{\l}(x+1)}{P_N^{\l}(x)} \; ;
\end{equation}
this polynomial  equals:
\begin{equation}
P_N^{\l}(x) = \prod\begin{Sb} 1 \leq j \leq N \\ j \not= i,i+1 \;;\; i\in I^{\l} \end{Sb} (x - 1 - \beta \l_j + j -\eta(\beta,N,M)) \; . \label{e: DrinfeldPoly}  
\end{equation}
The space $F^{\l}$ is also an eigenspace of the fermionic spin-$\frac{1}{2}$ Calogero-Sutherland Hamiltonian and of the degree operator:
\begin{eqnarray}
& & \\
\luN{H}_{CS}(\beta)\phi_{\tau}^{\l}(w) & = & E(\l;N,M)\phi_{\tau}^{\l}(w)\; ,\; \tau \in \tau(\l) \;; \nonumber \label{e: eigenvalue} \\
E(\l;N,M) & = & \sum_{i=1}^N(\beta \l_i^2 - 2i\l_i) + \nonumber \\ 
& & +((\beta+1)(2-N-M)+N)\sum_{i=1}^N\l_i + \nonumber \\ 
& & (\beta+1)\{\frac{N(N+M-2)^2}{4}-\nonumber \\ & & -\frac{1}{12}(N+M-2)(N+M-1)(N+M)\} \;; \nonumber \\
\sum_{i=1}^N D_i \phi_{\tau}^{\l}(w) & = & |\l| \phi_{\tau}^{\l}(w) \;. \label{e: Deigenvalue} 
\end{eqnarray}

Moreover, the decomposition of $\FS$ into spaces $ F^{\l}$, $ \l \in \Lambda_N^{(2)}$ is complete:
\begin{equation}
\FS = \oplus_{\l \in \Lambda_N^{(2)}} F^{\l} \; \label{e: FSDEC}
\end{equation}

Now expand the generating function $\Omw$ using the vectors $ \phi_{\tau}^{\l}(w) $ as a basis in $\FS$:
\begin{equation}
\Omw = \sum_{\l \in \Lambda_N^{(2)}}\sum_{\tau \in \tau(\l)}\omega^{(N),\l}_{M,\tau}\phi_{\tau}^{\l}(w) \; , \label{e: Omwexpansion} 
\end{equation}
and define a space $ \Omega^{(N),\l}_{M} \in \Om $ as follows: 
 
\begin{equation}
\Omega^{(N),\l}_{M} = \oplus_{\tau \in \tau(\l)} \cplx\omega^{(N),\l}_{M,\tau} \; , \label{e: Yirrep}
\end{equation}

The space $\Omega^{(N),\l}_{M}$ is an irreducible {\it highest weight} representation of $Y(\sll)$ defined by (\ref{e: Y0},\ref{e: Y1}), with the highest weight vector $\omega^{(N),\l}_{M,v^{\l}}$ and the Drinfel'd Polynomial $P_N^{\l}(x)$ (\ref{e: DrinfeldPoly}). It is also an eigenspace of the operators $H_{CS}(\alpha)$ and $L_0$: 
\begin{eqnarray}
H_{CS}(\alpha).\omega^{(N),\l}_{M,\tau}& = &  E(\l;N,M)\omega^{(N),\l}_{M,\tau}\;, \label{e: Hom} \\
L_0.\omega^{(N),\l}_{M,\tau} & = & h(\l;N,M)\omega^{(N),\l}_{M,\tau}\;, \quad\label{e: L0om} \tau \in \tau(\l)\; ,\\
h(\l;N,M) & = & |\l| + (\frac{M+N}{2} - 1)(\frac{M-N}{2}) \; . \nonumber
\end{eqnarray}
To be more precise, one  must verify, that none of the spaces $\Omega^{(N),\l}_{M}$ are equal to $0$ identically. 
This verification is done by computing the character $ F(q)^{(N)}_M $ = $ \text{tr}_{\Om}(q^{L_0})$ using the relation (\ref{e: L0om}), and assuming $\Omega^{(N),\l}_{M}$ $ \not\equiv 0 $,$  \l \in \Lambda_N^{(2)}$. The result of this computation is:
\begin{eqnarray}
F(q)^{(N)}_M & = & q^{\frac{M}{2}(\frac{M}{2} - 1)}\sum_{N^+ + N^- = N}\frac{q^{\frac{1}{4}(N^+ - N^-)^2}}{(N^+)_q (N^-)_q} \; , \label{e: OmChar}  \\ 
& & (r)_q = \prod_{n=1}^r(1 - q^n) \; .  \nonumber
\end{eqnarray}
The  expression above is seen to be a partition function of $N$ fermions of spin $\frac{1}{2}$. Alternatively the character can be computed directly -- without reference to the Yangian decomposition --  by using the hole basis in the space $\Om$ as written in the formula (\ref{e: Om}). The expression for the character so computed coincides with (\ref{e: OmChar}).  

Thus the Yangian decomposition of the space $\Om$ is found to be: 
\begin{equation}
\Om = \oplus_{\l \in \Lambda_N^{(2)}} \Omega^{(N),\l}_{M} \;. \label{e: Omexpansion}
\end{equation}
The Drinfel'd Polynomial of a representation $ \Omega^{(N),\l}_{M} $ is given by (\ref{e: DrinfeldPoly}); the eigenvalues of $H_{CS}(\alpha)$ and $L_0$ are specified by the equations (\ref{e: Hom}). 
It may be remarked that in the case of generic $\alpha$ ( or generic $\beta $ = $ -( \alpha + 2) $ ) which we are concerned with all the irreducible Yangian representations that appear in the decomposition (\ref{e: Omexpansion}) are isomorphic to tensor products of 2-dimensional evaluation representations of the Yangian \cite{CP}. 



\subsection{Decomposition of a space of semi-infinite wedges $F^{(M)}$ into irreducible representations of the Yangian}

To complete our discussion of the Yangian decomposition, we must find out which of the irreducible representations $\Omega^{(N),\l}_{M}$ that appear in the decomposition of the space $\Om$ (\ref{e: Omexpansion}) belong also to the space $\Omm$. This will make clear the irreducible content of each of the complements $\Om\setminus\Omm$  and, as can be seen from (\ref{e: dec1}) the irreducible Yangian content of the space $F^{(M)}$.   

Define a set $\Lambda_{N,0}^{(2)} $ as follows: 
\begin{equation}
\Lambda_{N,0}^{(2)} = \{ \l \in \Lambda_{N}^{(2)}| \l_{N-1} = \l_{N} = 0\}\;. \label{e: 0set}
\end{equation}

Let us demonstrate, that:
\begin{equation}\text{ if}\quad \text{  $\l$ $\in$ $\Lambda_{N}^{(2)}\setminus\Lambda_{N,0}^{(2)}$,} \quad  \text{ then $\Omega^{(N),\l}_{M}$ $\not\subset$ $\Omm$.}  \label{e: Exclusion}
\end{equation}
First, let $ \l_N > 0$. Then one has:
\begin{eqnarray}
\phi_{\tau}^{\l}(w) &  = & (w_1 w_2 \dots w_N) \phi_{\tau}^{\mu}(w) \;,\label{e: ff} \\
& & \text{where} \;  \mu_i = \l_i - 1 \; , \nonumber \\
& & \tau \in \tau(\l) = \tau(\mu) \; . \nonumber 
\end{eqnarray}
This is seen from  (\ref{e: Phi}) and  the relation \cite{M}: 
\begin{equation*}
J_{\l,N}^{(\beta)}(w)   =  (w_1 w_2 \dots w_N) J_{\mu,N}^{(\beta)}(w) \;.
\end{equation*}
The required property then follows from (\ref{e: ff}, \ref{e: OmwInclusion}).   

Now, let $ \l_N = 0$ but $ \l_{N-1} > 0$. In this case the Drinfel'd Polynomial (\ref{e: DrinfeldPoly}) of the representation  $\Omega^{(N),\l}_{M}$ contains the root:
\begin{equation*}
 \gamma(\beta,N,M) = 1+\beta\l_i - i + \eta(\beta,N,M) |_{i=N}  =     (\beta + 2)(1 - \frac{N+M}{2}) \; ,\end{equation*}
which can never appear as a root of a Drinfel'd Polynomial associated with any of the irreducible Yangian sub-representations in $\Omega^{(N-2)}_{M}$ as long as $\beta$ is generic. Hence follows the property (\ref{e: Exclusion}).  

A computation of the character of the space :
\begin{equation}
   \oplus_{\l \in \Lambda_{N,0}^{(2)}} \Omega^{(N),\l}_{M}  \; \; \subset \; \;  \Om \; ;
\end{equation}
shows, that this character is equal to the character of the space $\Omm$. This, and (\ref{e: Exclusion}) gives the desired decomposition of the space $\Om\setminus\Omm$: 

\begin{equation}
\Om\setminus\Omm = \oplus_{\l \in \Lambda_N^{(2)}\setminus\Lambda_{N,0}^{(2)}} \Omega^{(N),\l}_{M} \;. \label{e: quot}
\end{equation}
The last equation together with ($\ref{e: dec1}$), constitute the irreducible decomposition of the space $F^{(M)}$ with respect to the Yangian action defined by (\ref{e: Y0},\ref{e: Y1}):
\begin{equation}
F^{(M)} = \bigoplus\begin{Sb} N \geq 0 \\ N+M = \text{even} \end{Sb} \left( \oplus_{\l \in \Lambda_N^{(2)}\setminus\Lambda_{N,0}^{(2)}} \Omega^{(N),\l}_{M} \right) \; . \label{e: Maindec}
\end{equation}
One can reformulate the result of this decomposition in a somewhat different way. Let $\Lambda^{(2)}$ be the set of all partitions which have no more than two equal parts of any given value ( set of all Young diagrams which have no more than two rows of any given length ). One has:
\begin{equation}
\Lambda^{(2)} = \bigcup\begin{Sb} N \geq 0 \\ N+M = \text{even} \end{Sb} \Lambda^{(2)}_N\setminus\Lambda^{(2)}_{N,0}\;.\label{e: Young}
\end{equation} 
For $\l$ $\in$ $\Lambda^{(2)}$ and $M$ even(odd) define $\Omega^{\l}_M$ as follows: 
\begin{equation*}
\Omega^{\l}_M = \begin{cases} \Omega^{(N=l(\l)),\l}_M\; & l(\l) = \text{even(odd),  } \\ \Omega^{(N=l(\l)+1),\l}_M\; & l(\l) = \text{odd(even). }\end{cases}
\end{equation*}
The decomposition (\ref{e: Maindec}) of the space $F^{(M)}$ with $M$ even(odd) then can be written as:
\begin{equation*}
F^{(M)} = \oplus_{\l \in \Lambda^{(2)}} \Omega^{\l}_M \;,
\end{equation*}
where $\Omega^{\l}$ is a highest-weight irreducible representation of the Yangian with the Drinfel'd Polynomial $P^{\l}(x)$ :
\begin{equation}
P^{\l}(x) = \begin{cases} P^{\l}_{l(\l)}(x)\; , & l(\l) = \text{even(odd) ,} \\
 P^{\l}_{l(\l)}(x)(x - \gamma(\beta,\l,M)) \; , & l(\l) = \text{odd(even) .}\end{cases}
\end{equation}
Here $P^{\l}_{l(\l)}(x)$ = $ P^{\l}_{N=l(\l)}(x)$ is the polynomial defined in (\ref{e: DrinfeldPoly}), and 
\begin{equation*}
\gamma(\beta,\l,M) = (1 + \frac{\beta}{2})(1 - l(\l) - M ) \;.
\end{equation*}

Finally, one can compute the character $ F(q)_M $ = $ \text{tr}_{F^{(M)}}(q^{L_0})$  of the space of semi-infinite wedges $F^{(M)}$ using the decomposition (\ref{e: Maindec}): 

\begin{eqnarray}
& &   \\
F(q)_M & = & \sum\begin{Sb} N \geq 0 \\ N+M = \text{even} \end{Sb} F(q)^{(N)}_M -  F(q)^{(N-2)}_M \; = \nonumber \label{e: character} \\
& =  & \lim\begin{Sb} N \rightarrow \infty \\ N+M = \text{even} \end{Sb} F(q)^{(N)}_M \; = \nonumber \\
& = & \frac{q^{\frac{M}{2}(\frac{M}{2} - 1)}}{(\infty)_q^2} \sum_{p\in \zint} q^{(p + \frac{\ep}{2})^2} \; , \; \ep = 0 (1) \text{when $M$ is even ( odd ).}\nonumber
\end{eqnarray}
The last expression above is the known character formula for the space $F^{(M)}$ which can be obtained, for example, from the decomposition (\ref{e: decomp}).

\section{Concluding remarks}

We have seen, that the language of the wedge product formalism is well-suited for derivation of the low-energy, conformal field theory limit of the fermionic spin Calogero-Sutherland Model. One can think of several other situations where the approach based on wedges might be fruitful. 
First of all the generalization to the $\frak{s}\frak{l}_n$ case seems to be quite straightforward. Another generalization - to the $q$-deformed situation may be approached by using the $q$-wedges recently introduced in \cite{St} and \cite{KMS}. 
This should provide a way to describe a low-energy limit of Ruijsenaars models with spin \cite{BGHP}, \cite{Konno}; and connect with the results of \cite{JKKMP}.

\mbox{}

{\large {\bf Acknowledgments }}\\
I wish to thank Professors T.Miwa and M.Kashiwara for their kind interest in this work and many stimulating discussions.

\end{document}